\documentclass[conference]{IEEEtran}

\usepackage[latin9]{inputenc}
\usepackage{amsmath}
\usepackage{graphicx}
\usepackage{subfigure}
\usepackage{bm}
\usepackage{hyperref}
\usepackage{textcomp}

\makeatletter

\begin{document}

\title{A Classification Engine for \\Image Ballistics of Social Data}

%\author{Oliver Giudice, Antonino Paratore, Marco Moltisanti, Sebastiano Battiato}

% author names and affiliations
% use a multiple column layout for up to three different
% affiliations
\author{\IEEEauthorblockN{Oliver Giudice\IEEEauthorrefmark{1}\IEEEauthorrefmark{2},
Antonino Paratore\IEEEauthorrefmark{2},
Marco Moltisanti\IEEEauthorrefmark{1} and
Sebastiano Battiato\IEEEauthorrefmark{1}\IEEEauthorrefmark{2}}
\IEEEauthorblockA{\IEEEauthorrefmark{1}Dipartimento di Matematica e Informatica, University of Catania,\\ Email: giudice,moltisanti,battiato@dmi.unict.it}
\IEEEauthorblockA{\IEEEauthorrefmark{2}iCTLab s.r.l. - Spinoff University of Catania\\
mail: antonino.paratore@ictlab.srl}}

% The paper headers
%\markboth{Journal of \LaTeX\ Class Files,~Vol.~13, No.~9, September~2014}%
%{Shell \MakeLowercase{\textit{et al.}}: Bare Demo of IEEEtran.cls for Journals}
% The only time the second header will appear is for the odd numbered pages
% after the title page when using the twoside option.
% 
% *** Note that you probably will NOT want to include the author's ***
% *** name in the headers of peer review papers.                   ***
% You can use \ifCLASSOPTIONpeerreview for conditional compilation here if
% you desire.

% make the title area
\maketitle

% As a general rule, do not put math, special symbols or citations
% in the abstract or keywords.
\begin{abstract}
Image Forensics has already achieved great results for the source camera identification task on images. Standard approaches for data coming from Social Network Platforms cannot be applied due to different processes involved (e.g., scaling, compression, etc.). Over 1 billion images are shared each day on the Internet and obtaining information about their history from the moment they were acquired could be exploited for investigation purposes.
In this paper, a classification engine for the reconstruction of the history of an image, is presented. Specifically, exploiting K-NN and decision trees classifiers and a-priori knowledge acquired through image analysis, we propose an automatic approach that can understand which Social Network Platform has processed an image and the software application used to perform the image upload. The engine makes use of proper alterations introduced by each platform as features. Results, in terms of global accuracy on a dataset of 2720 images, confirm the effectiveness of the proposed strategy.
\end{abstract}

% Note that keywords are not normally used for peerreview papers.

% For peer review papers, you can put extra information on the cover
% page as needed:
% \ifCLASSOPTIONpeerreview
% \begin{center} \bfseries EDICS Category: 3-BBND \end{center}
% \fi
%
% For peerreview papers, this IEEEtran command inserts a page break and
% creates the second title. It will be ignored for other modes.
\IEEEpeerreviewmaketitle

\section{Introduction}
% * <giudice@dmi.unict.it> 2016-07-16T15:51:35.807Z:
%
% dettagliare meglio il problema: quali sono le problematiche legate ai social che non permettono di fare image forensics come sempre? Quali ulteriori problemi creano?
%
% ^.
Image Forensics traditionally refers to a number of different tasks on digital images aiming at producing evidence on the authenticity and integrity of data (e.g., forgery detection) and on the identification of the acquisition device (camera identification) \cite{piva2013},\cite{stemm2013}.
Digital images are continuously altered starting from the moment they were acquired. Most of the time, these alterations are made by users with precise malicious intents. Typical tamperings are the removal or the insertion of an object in an image, the cropping of an undesired portion of a picture, or the application of particular filters to modify or mask sensible parts (e.g., faces in pedo-pornographic photos). When the tampering is not clearly visible, the problem of detecting it becomes obviously challenging.
To solve the forgery detection task, some approaches stand above the others: a group of them looks at the structure of the file (e.g., JPEG blocking artifacts analysis \cite{bruna2011}, \cite{luo2007}, hash functions \cite{battiato2012}, JPEG headers analysis \cite{kee2011}, thumbnails \cite{kee2010} and EXIF analysis \cite{gloe2012}, etc.); others try to identify the device that acquired the image by making use of PRNU patterns (\cite{chen2012},\cite{dirik2014}), or focus on statistical analysis of the DCT coefficients (\cite{redi2011,galvan2014,Battiato2009}). A voting approach has been used for the same purpose in \cite{battiato2012}.
Another important task for Image Forensics is finding the camera device that acquired the image. Some in-depth studies (\cite{farid2008digital} \cite{Kornblum2008}) showed that it is possible to coarsely solve the camera identification task, using the DCT coefficients as a feature. 
All these works make clear the importance of the JPEG pipeline in retrieving information about the history of an image.

Nowadays Social Networks allow their users to upload and share large amounts of images: just on \textit{Facebook} about 1 billion images are shared every day. 
What happens when a picture is shared on a social platform? How does the upload process affect the JPEG elements of the image?
A Social Network is yet but another piece of software that alters images for bandwidth, storage and layout reasons. These alterations have been proved to make state-of-the-art approaches for camera identification less precise and reliable.
Recent studies (\cite{Moltisanti2015},\cite{castiglione2011}) have shown that, although the platform heavily modifies an image, this processing leaves a sort of fingerprint on the image itself. This evidence can be exploited to understand if the image has been actually uploaded to the \textit{Facebook} platform. State of the art studies, regarding Social Platforms, focus on the analysis of alterations on images and do not propose a method to solve the image ballistics task. Moreover, they focus on too few Social Network platforms making their works not general enough. Hence, enlarging the analysis to further Social Network Services (SNSs), the reconstruction of the history of an image becomes a difficult task.
To understand how SNSs process images, a dataset of images from different camera devices was collected, under controlled conditions. We selected ten SNSs through which we processed the collected images by mean of an upload and download process. By doing this, a dataset of images has been obtained, in order to identify any alterations on JPEG elements.
The main discovery of our study was that alterations observed are platform dependent (server-side) but also related to the application carrying out the upload (client-side). This evidence can be fundamental for investigation purposes to understand not only the provenience of an image, but also if it has been uploaded from a given device (e.g., Android, iOS). 
All the observed alterations allowed to build an automatic classifier, based on two K-NN classifiers and a decision tree fitted on the built dataset. Starting from an input image, the proposed approach can predict the SNS that processed the image and the client application through which the image has been uploaded.
The remainder of the paper is structured as follows: in Sec. \ref{sec:construction}, we describe how the dataset has been built, which social platforms have been considered and what kind of upload methods have been used; in Sec. \ref{sec:analysis}, an in-depth analysis on dataset images is reported in order to find alterations that can be coded into a fingerprint for a SNS processing; in Sec. \ref{sec:ballistics}, our approach for image ballistics on social image data is presented with the obtained classification results. Finally, conclusions and reasoning about possible future works on the topic are discussed.

% needed in second column of first page if using \IEEEpubid
%\IEEEpubidadjcol

\section{A Dataset of Social Imagery}
\label{sec:construction}
\begin{table*}
\centering
\caption{Devices used to carry out image collection. For each device the corresponding Low Quality (LQ) and High Quality (HQ) resolutions are reported.}
\begin{tabular}{ |l|l|l|l| }
  \hline
  \textbf{Model} & \textbf{Device Type} & \textbf{Low Resolution} & \textbf{High Resolution} \\
  \hline
  \textbf{Canon Eos 650D} & Dedicated device & 720x480 & 5184x3456 \\
  \textbf{QUMOX SJ4000} & Dedicated device & 640x480 & 4032x3024 \\
  \textbf{Sony Powershot A2300} & Dedicated device & 640x480 & 4608x3456 \\
  \textbf{Samsung Galaxy Note 3 Neo} & Android 4 Phone & 640x480 & 3264x2448 \\
  \textbf{HTC Desire 526g} & Android 5 Phone &  640x480 & 3264x2448 \\
  \textbf{Huawei G Play Mini} & Android 6 Phone & 640x480 & 4208x3120 \\
  \textbf{iPhone 5} & iOS 6 Phone & 640x480 & 2448x3264 \\
  \textbf{iPad mini 2} & iOS 8 Pad & 640x480 & 800x600 \\
  \hline
\end{tabular}
\label{tab:devices}
\end{table*}
The alterations introduced on images by SNS can be thought as a unique fingerprint left by the SNS. The aim of our study is to discover those fingerprints by analyzing the behavior of the most popular SNSs that allow image sharing. Hence, 10 platforms have been selected. First of all, \textit{Facebook} (http://www.facebook.com) and \textit{Google+} (http://plus.google.com) were taken into account as being the two most popular platforms where users can share their statuses and multimedia content to a network of friends. \textit{Twitter} (http://www.twitter.com) and \textit{Tumblr} (http://www.tumblr.com) were considered as being representative of the micro-blogging concept. We included also \textit{Flickr} (https://www.flickr.com) and \textit{Instagram} (https://www.instagram.com) as platforms focused on sharing high quality artistic photos with capabilities of image editing and filtering. \textit{Imgur} (http://www.imgur.com) and \textit{Tinypic} (htto://www.tinypic.com) were also taken into consideration even if they are not properly SNSs but are very popular platforms for image sharing: users usually link images hosted on them from forums and web sites all over the Internet. Finally \textit{WhatsApp} (http://www.whatsapp.com) and \textit{Telegram} (http://www.telegram.org) were also selected as being the two most popular mobile messaging platforms that, by allowing users to create chat groups, are another big place for image sharing on the Internet. Specifically, the last two services are often involved in forensic investigations.

To discover how SNSs process images, we collected a set of photos with the camera devices listed in Table \ref{tab:devices}.
Images were acquired representing three different types of scenes: outdoor scenes with buildings (artificial environment), outdoor scenes without buildings (natural environment) and indoor scenes.
When taking a picture, we captured two versions: a High Quality (HQ) photo at the maximum resolution allowed by the device, and a Low Quality (LQ) photo (see also Table \ref{tab:devices}). Capturing images in this way, a dataset with a good variability in terms of contents and resolutions was obtained.

Images collected so far were uploaded to each of the considered platforms with two different methods: with a web browser, and with iOS and Android native apps. No further discrimination is needed for web browsers because we observed that alterations are not browser-dependent.
Each download was performed by searching for the image file URL in the HTML code of the page showing the image itself. At the end of this phase 2400 images were properly collected.

The second upload method was carried out with iOS and Android native apps of each social platform, except for \textit{Tinypic} that do not possess an official app in stores. Moreover, the upload has been done by choosing images in two ways: by searching in the gallery for a previously acquired image (images from local gallery) and by acquiring the image with the camera app embedded in the app itself (embedded camera app). After uploading all images as described above, all of them were downloaded through the "URL searching technique" previously described. 320 more images processed through 8 platforms were thus obtained.
All uploads were performed with default settings.

The overall dataset consists of 2720 images in JPEG format and it is available at the following web address \url{http://iplab.dmi.unict.it/DigitalForensics/social_image_forensics/}.

\section{Dataset Analysis}
\label{sec:analysis}
The main aim of our work is to find a fingerprint left by SNSs on JPEG structure elements, after an upload/download process, in order to build a classifier for image ballistics.
To achieve this goal, all information contained in the JPEG file specification has been analyzed: image filename, image size, meta-data and JPEG compression information. We observed that each upload/download process through the considered SNSs produces different alterations among the above-mentioned elements that could be taken into account as fingerprints of the process itself. Details of these alterations will be described in the following Subsections.

\subsection{Image Filename Alterations}
\label{sec:filename}

\begin{table*}
\centering
\caption{Renaming scheme for an uploaded image with original filename IMG\_2641.jpg. The  new file name for each platform is reported (Image IDs are marked in bold).}
\begin{tabular}{ |l|l|l|l| }
  \hline
  \textbf{Social} & \textbf{Rename (image ID in bold)} & \textbf{Image Lookup} & \textbf{Other information}\\
  \hline
  \textit{Facebook} & 11008414\_\textbf{746657488782610}\_8508378989307666639\_n.jpg & YES & Upload resolution \\
  \textit{Flickr} & 26742193671\_\textbf{8a63f10c85}\_h.jpg & YES & Download resolution (h=1600)\\
  \textit{Tumblr} & tumblr\_\textbf{o3q9ghRCRh1vnf44lo9}\_1280.jpg & YES & Download resolution (1280)\\
  \textit{Imgur} & 04 - \textbf{Dw0KXG2}.jpg & YES & \\
  \textit{Twitter} & \textbf{CdqCPQ-WAAAzrHI.jpg} & YES & \\
  \textit{WhatsApp} & IMG-20160314-WA0038.jpg & NO & Receiving Date (2016-03-14)\\
  \textit{Tinypic} & 1zqdirm.jpg & NO & \\
  \textit{Instagram} & 1689555\_\textbf{169215806798447}\_744040439\_n.jpg & YES & Upload Resolution \\
  \textit{Telegram} & 422114602\_5593965449613038107.jpg & NO & \\
  \hline
\end{tabular}
\label{tab:rename}
\end{table*}

\begin{table*}
\centering
\caption{Alterations on JPEG files. The EXIF column reports how JPEG meta-data are edited: maintained, modified or deleted. The File Size column reports if a resize is applied and the corresponding conditions. The JPEG compression column reports if a new JPEG compression is carried out and the corresponding conditions (if any).}
\begin{tabular}{|c|c|c|c|c|c|c|}
\cline{1-7}
 
 \textbf{Social} & \multicolumn{2}{c|}{\textbf{EXIF}} & \multicolumn{2}{c|}{\textbf{File Size}} & \multicolumn{2}{c|}{\textbf{JPEG Compression}} \\
\hline 
	{} & \textbf{Camera Data} & \textbf{Other Data} & \textbf{Resize} & \textbf{Resize Condition}  & \textbf{Re-Compression} & \textbf{Re-Compression Condition} \\ 
\hline
\textit{Facebook} & Delete & Delete & Yes & LQ: $M > 960$ HQ: $M > 2048$ & Yes & Always\\
\textit{Google+}  & Maintain & Maintain/Edit & Yes & $M > 2048$ & Yes & $M > 2048$\\
\textit{Flickr}   & Delete & Maintain/Edit & Yes & Depends on options & Yes & Depends on options \\
\textit{Tumblr}   & Maintain & Maintain/Edit & Yes & $M > 1280$ & Yes & $M > 1280$\\
\textit{Imgur}	& Delete & Delete & No & Never & Yes & Image Size (MB) $>$ 5.45 MB \\
\textit{Twitter} & Delete & Delete & Yes & $M > 2048$ & Yes & Always \\
whatsApp & Delete & Delete & Yes & $M > 1600$ & Yes & Always \\
\textit{Tinypic} & Maintain & Maintain/Edit & Yes & $M > 1600$ & Yes & $M > 1600$\\
\textit{Instagram} & Delete & Delete & Yes & $M > 1080$ & Yes & Always\\
\textit{Telegram}  & Delete & Delete & Yes & $M > 2560$ & Yes & Always\\
\hline

\hline

\end{tabular}
\label{tab:resume}
\end{table*}

The analysis of the filename of an image and the comparison with known patterns during an investigation on storage devices can provide information about the platform from which it could be downloaded and the date when it was uploaded.
For this reason, we first evaluated if and how each platform modifies the file name. We observed that all platforms except \textit{Google+} do a rename. 

As an example, in Table \ref{tab:rename} the new names for an uploaded file with name "IMG\_2641.jpg" are reported. The column "image lookup" describes the presence into the new filename of an ID useful to reconstruct an URL that points to the web location where the image file is stored. 

\textit{Facebook}, \textit{Flickr}, \textit{Tumblr} and \textit{Instagram} use the image ID and the platform public API (e.g., Graph for \textit{Facebook}) to build the corresponding URL.
\textit{Twitter} and \textit{Imgur} allow finding the image on the respective platform by navigating to the URL:
\begin{itemize}
\item https://pbs.twimg.com/media/$<$IMAGE ID$>$ for \textit{Twitter};
\item http://imgur.com/$<$IMAGE ID$>$ for \textit{Imgur};
\end{itemize}
The other platforms do not present an image ID. 

%Moreover there could be also other useful information like the receiving date (for Whatsapp) and the image resolution (\textit{Facebook}, \textit{Flickr}, \textit{Tumblr} and \textit{Instagram}) coded into image filenames.

%File naming alone can solve the problem of identifying the SNS, but it is weak to be used in digital investigations, because filenames can be easily modified by the user.
%For instance on \textit{Instagram}, the easiest way to download an image is by right-clicking it and choose "save as". By doing this, the browser instantly modifies the name of the image that is going to be downloaded.

\subsection{Image Size Alterations}
A stronger evidence than file naming is the resize of the uploaded images on each platform. A fine-grained test was performed by using synthetic images derived from our dataset and resized at different scales. 

On most platforms, resizing is applied if and only if the input image matches certain conditions. This condition is linked to the length in pixels of the longest side $M$ of the original image, where $M = max(width, height)$. If $M$ is greater than a threshold, a resizing algorithm is applied and the resulting image has its longest size equal to the threshold.
In Table \ref{tab:resume}, such conditions and the corresponding thresholds for each platform are reported. \textit{Tumblr} does not rescale uploaded images, while in \textit{Flickr} the threshold is set by the user. When the images are resized, the longest side will be set to a fixed value that identifies, in some sense, the platform that made the operation (see Table \ref{tab:resume}). Let note that, some of the considered platforms use the same threshold value and it is subject to changes over time (for example, during our experiments, the threshold for \textit{Twitter} changed from 1024 to 2048).
%If a resizing algorithm is applied the obtained image has its longest side with a value corresponding to the platform that made the operation. Unfortunately some platforms share the same values and this values too can vary through time. For example during our experiments the threshold for twitter changed from 1024 to 2048 pixels.

\subsection{Meta-data Alterations}

\begin{figure*}
\caption{Classification scheme for Image Ballistics in the era of Social Network Services. The proposed approach encodes JPEG information from an input image into a feature vector. The obtained feature vector is evaluated through an Anomaly Detector that filters out images not  processed by a SNS. If the input image is not an anomaly, the feature vector goes through other two classifiers: a SNS Classifier and an Upload Client Classifier. The output of the SNS Classifier is further processed through a SNS Consistency Test that checks if the features of the input image and the predicted SNS are consistent to re-compression and resizing conditions. The final output depends on this last stage: if all features are compatible with the classified SNS then the obtained prediction, joined with upload client prediction, is outputted. Otherwise the consistency test is repeated, for the next most probable predicted SNS, until it is satisfied or it stalls on the same predicted platform. In this case the overall output will be "Not Sure".}
\label{fig:classification}
\centering
\includegraphics[width=1.8\columnwidth] {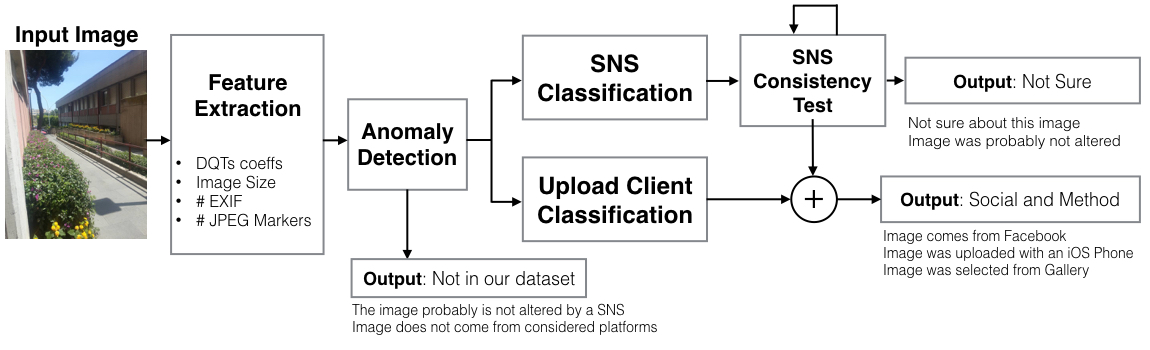}
\end{figure*}

The best evidence to obtain information, for investigation purposes, are meta-data embedded in JPEG files. These meta-data are technically known as EXIF and can store information like the device that acquired an image, the date and time of acquisition and also the GPS coordinates. For our purposes, EXIF data were divided into two categories: ``camera data'' which contains all those key-valued that allow to identifying the device that acquired the image and ``other data'' for every other information.

In Table \ref{tab:resume}, the results of the analysis on EXIF data are resumed for each platform. In particular, it is reported if "camera data" and "other data" are deleted, maintained or just edited throughout the processing. Unfortunately, most of the SNSs delete all meta-data, specifically those related to camera data.

\subsection{Image JPEG Compression Alterations}
\label{sec:compression_alt}

The images considered in our dataset are all encoded in JPEG format, both the original versions and the downloaded ones. Thus, an analysis on how the SNS processing affects the JPEG compression has been carried out. We focused on the Discrete Quantization Tables (DQTs) used for JPEG compression (extracted by DJPEG: an open source tool part of libjpeg project \cite{djpeg}).

Considering how platforms affect DQTs, it is possible to divide them into two categories:
\begin{itemize}
\item Platforms that always re-compress images (\textit{Facebook}, \textit{Twitter}, \textit{Telegram}, \textit{WhatsApp}, \textit{Instagram});
\item Platforms that re-compress images at a given condition (\textit{Google+}, \textit{Tumblr}, \textit{Tinypic}, \textit{Imgur}).
\end{itemize}

The compression follows the same rules we described for resizing. In fact, a threshold-based evaluation is performed on the longest image side and, if it is bigger than the threshold, the image is compressed using a DQT that will be different from the original one. This is not true for all the considered platforms; \textit{Flickr} allows the user to choose the threshold (if any), while on \textit{Imgur} the threshold is fixed in terms of size in MegaBytes; specifically, if the input image size is greater than 5.45MB, than the re-compression is performed, otherwise nothing happens (see also Table \ref{tab:resume}).
%In a same way as described for the resizing algorithm, the condition for re-compression is: if the size in pixels of the longest side of input image $M$ is smaller than a threshold, that is typical of the platform, then a new JPEG compression is applied and the obtained image will have different DQTs than the input one, in terms of their coefficients.

%Some additional words are needed for \textit{Flickr} and \textit{Imgur}. As regards \textit{Flickr} as described previously for resizing, the condition is set by the user. While for \textit{Imgur}, the condition is based on size in MegaBytes (MB): if the input image size is greater than 5.45MB than the re-compression is performed, otherwise nothing is done.
%Some additional words are needed for \textit{Imgur} for which the condition is based on size in MegaBytes (MB): if the input image size is greater than 5.45MB than the re-compression is performed, otherwise nothing is done.

\section{Image Ballistics of Social Data}
\label{sec:ballistics}
Starting from the results of the analysis reported in previous Sections, regarding the alterations on JPEG elements of processed images, it is possible to assess that such alterations bring pieces of information about the history of the image but they could be insufficient, if considered alone, for investigation purposes. Hence, we represent all the observed alterations into a set of features to be used as input for an automatic classifier. 
The following elements are then embedded into proper numerical features:

%For this reason by considering the alterations observed altogether it is possible to identify them as a fingerprint to be used in an automatic classifier. In order to do this we coded the alterations observed into the following numeric features:
\begin{itemize}
\item The DQTs coefficients divided in 64 coefficients for the Chrominance table and 64 for the Luminance one, which represent the JPEG compression alterations;
\item Image size (width and height in pixels), which brings information about size alterations;
\item Number and typology of EXIF data (key-value couples), which describes meta-data alterations;
\item Number of markers in JPEG files as defined in \cite{miano1999compressed}.
\end{itemize}

The listed features were chosen in order to represent each kind of alteration described in previous Sections. 

The Quality Factor (QF) was not considered among them, as it was done in \cite{Moltisanti2015}, for being dependent on DQTs coefficients and thus not bringing any new useful information. In particular, QF does not have a unique method to be computed and this can be be a source of error for classification purposes. 

PRNU was also not taken into consideration among our features, because, as already mentioned, the heavy processing done on images by SNSs destroys/modifies any information coming from the sensor which acquired the image.

Given the features listed before and the image dataset described so far, a correspondence between features and the SNS has been established. This is particularly true for platforms that always operate a re-compression and heavily alter images. Starting from this correspondence, an automatic classification approach for image ballistics was built. Given an input image, the proposed method allows knowing not only from which platform it comes from, but also which client application was used to upload the image (browser web application, iOS native app or Android native app). Moreover, for images uploaded from iOS and Android native apps, the proposed approach is able to differentiate between images taken from the camera application embedded into the native apps or images chosen through gallery selection. This demonstrates that fingerprints observable on images are left both by SNSs and the client applications carrying out the upload.

\subsection{Implementing image ballistics: a classification engine}
%%% NEW VERSION STARTING HERE
Given a JPEG image $I$, our objectives are to define:
\begin{enumerate}
\item if there is a compatibility between the non-related JPEG elements of $I$ (i.e. filename, EXIF data) and the processing pipeline of SNSs;
\item if there is a compatibility between the JPEG elements of $I$ and the processing pipeline of SNSs;
\item which SNS is compatible with the JPEG elements of the image, with a certain degree of confidence, and what is the uploading source in terms of operating system (OS) and application.
\end{enumerate}

We represent each image $I$ as a 44-dimensional vector 
\begin{equation}
\label{eq:feature_vector}
\boldsymbol{v} = \{w,h,|E|,m,l_j,c_k\},
\end{equation}
where
\begin{itemize}
	\item $w \times h$ is the size in pixels of $I$;
    \item $E = \left\lbrace key, value \right\rbrace$ is an associative array containing the EXIF metadata, thus $\|E\|$ is the number of metadata found in the structure of $I$;
    \item $m$ is the number of JPEG markers in $I$;
    \item $l_j, j=0,\ldots,31$ are the first 32 coefficients of the luminance quantization table;
    \item $c_k, k=0,\ldots,7$ are the first 8 coefficients of the chrominance quantization table.
\end{itemize}

Moreover, we define $fn\left(I\right)$ as the filename of the image $I$.
%Let $I$ of size $w \times h$ pixels be the JPEG image, and let $m$ be the number of markers in the JPEG file structure. We refer to the EXIF data as an associative array $E=\{key,value\}$.
%, and presenting a JPEG file structure composed of $m$ markers; we propose a classification engine to solve the image ballistics task for $I$.

At the first stage, we consider $fn\left(I\right)$ and $E$. If there is a matching between $fn\left(I\right)$ and the renaming patterns observed in Section \ref{sec:filename}, our approach confirms the compatibility between $I$ and the SNS with the matched pattern. Also, $E$ is taken into account, looking for the ``Exif.Image.UniqueCameraModel'' key. If it is set, then our system returns that value.

Thus, the whole dataset representation is
\[
\boldsymbol{V} = \left\lbrace \boldsymbol{v}_1, \boldsymbol{v}_1, \ldots, \boldsymbol{v}_N \right\rbrace
\]

where $N$ is the total number of images. In order to train the SNS and Upload Scenario classifiers, we augment this representation with the corresponding labels. Thus, the final representation for a generic image $I_i$ is
\[
\boldsymbol{l}_i = \left\lbrace \boldsymbol{v}_i, sns_i, uc_i, sm_i \right\rbrace
\]

where $sns_i$ is the SNS, $uc_i$ is the client application and $sm_i$ is the image selection method.

%At first, the image filename is taken into account. If it corresponds to any of the SNS rename patterns observed in Section \ref{sec:filename}, our approach outputs that the input image is compatible with the corresponding SNS. At this stage EXIF data are also taken into account with the following condition: if the camera model is set, it is outputted, otherwise nothing can be said. 

%The actual classification engine proposed starts with the encoding of the features of $I$ in a feature vector:

%\begin{equation}
%\label{eq:feature_vector}
%\boldsymbol{v} = \{w,h,e,m,l_0,...,l_31,c_0,...,c_8\},
%\end{equation}

%where $w$ and $h$ are respectively the width and height of $I$ in pixels, $e = |E|$, $m$ is the number of JPEG markers in $I$, finally $l_i$ and $c_j$ are the most discriminative coefficients respectively for the Luminance and Chrominancs DQTs. Thus $v$ is a 44-dimensional feature vector of positive integers.

Our classifier performs a two-steps analysis. First, we implement an Anomaly Detector to exclude the images that have not been processed by SNSs, then we run in parallel a K-NN Classifier and a Decision Tree \cite{quinlan1986induction} to asses respectively the SNS of origin and the uploading scenario (OS + application).

Given the representations $\boldsymbol{v}_{I_1}$ of an image $I_1$ and $\boldsymbol{v}_{I_2}$ of an image $I_2$, we define the cosine distance between $\boldsymbol{v}_{I_1}$ and $\boldsymbol{v}_{I_2}$

\begin{equation}
d(\boldsymbol{v_1},\boldsymbol{v_2}) = \frac{\boldsymbol{v_1} \cdot \boldsymbol{v_2}}{|\boldsymbol{v_1}| |\boldsymbol{v_2}|}
\end{equation}

as a measure of similarity between $I_1$ and $I_2$. Therefore, it is possible to build a distance matrix $\boldsymbol{D}$ of size $N \times N$ where the element $d_{ij}$ is equal to the distance between the images $I_i$ and $I_j$. We will refer to the $r-$th row of this matrix as $\boldsymbol{D}_r$ and to the $c-$th column as $\boldsymbol{D}^c$. 
It is important to note that $\forall\; I_i,I_j,\;0 \leq d(\boldsymbol{v_i},\boldsymbol{v_j}) \leq 1$, and specifically, the more is the similarity, the more the distance will be closer to 1. Exploiting this property, we define the Anomaly Detector as 

\begin{equation}
\label{eq:anomaly_condition}
a\left( \boldsymbol{v}_i, \boldsymbol{D} \right) = 
\left\lbrace
  \begin{array}{ll}
      %d\left(\boldsymbol{v}_i, \min\boldsymbol{D}^i\right) & if\;\sum\limits_{j=1}^K d_{ij} < T \\
      \left(\boldsymbol{v}_i,i\right) & if\;\sum\limits_{j=1}^K d_{ij} < T \\
      \textnormal{not processed} &  otherwise
  \end{array} 
\right.
\end{equation}

where $T \in [0,K]$ is defined as the Anomaly Threshold. In other words, since the more two images are similar, the more their distance will be closer to 1, we make sure that at least $\lfloor K\rfloor$ samples in our dataset are similar to the query image representation. 
Then, when $a\left( \boldsymbol{v}_i, \boldsymbol{D} \right) = 0$, the representation is far apart the samples, and we can state that probably the image has not been processed.

The output of $a$ is then used as input by 1-NN (\ref{eq:knn}) and Decision Tree algorithms \cite{quinlan1986induction}.

\begin{equation}
\label{eq:knn}
knn\left(\boldsymbol{v}_i, i\right) = sns_j \; | \; d_{ij} = \min\boldsymbol{D}^i
\end{equation}

\begin{equation}
\label{eq:decision_tree}
dt\left(\boldsymbol{v}_i, i \right) = (uc_j,sm_j)
\end{equation}
where  $uc_j$ and $sm_j$ are the leaves obtained following the path with $\boldsymbol{v}_i$ as input. Hence, the classification scheme, shown in Figure \ref{fig:classification}, can be formalized as follows
\begin{equation}
\label{eq:classifier}
C(\boldsymbol{v}_i,\boldsymbol{D}) = knn\left(a\left(\boldsymbol{v}_i, \boldsymbol{D} \right)\right) \oplus dt\left(a\left(\boldsymbol{v}_i, \boldsymbol{D} \right)\right)
\end{equation}

1-NN algorithm looks for the closest sample in the dataset, and assigns the same SNS to the query image. A Decision tree (Eq. \ref{eq:decision_tree}) builds classification in the form of a tree structure. It breaks down a dataset into smaller and smaller subsets while at the same time an associated decision tree is incrementally developed. The final result is a tree with decision nodes. The algorithm used for building the decision tree is the ID3 \cite{quinlan1986induction} which employs a top-down, greedy search through the space of possible branches with no backtracking. ID3 uses Entropy to construct a decision tree by evaluating $\boldsymbol{v} \in \boldsymbol{V}$.

Finally, the output of the 1-NN Classifier $sns_j$ is processed through a SNS Consistency Test. 
Let be $S = \{sns_1,\ldots,sns_n\}$ the set of SNSs that operates a re-compression at the condition $max(w,h) > C_{sns_i}$ where $C_{sns_i}$ is the conditional threshold for the $i-$th SNS and $w$ and $h$ as listed in Table \ref{tab:resume}.

Given that $sns_j \in S$, if $max(w,h) < C_{sns_j}$ it is an anomaly. The test is then repeated for the next most probable prediction from the SNS Classifier until the corresponding condition is satisfied or the loop stalls on the same SNS prediction. In this last case, the result of the classification is "not sure"; otherwise, a SNS prediction is reached and outputted ($sns_j$) with the predicted upload client application ($uc_j$) and image selection method ($sm_o$).

Figure \ref{fig:classification} shows a schematic representation of the proposed classification engine.

\subsection{Classification Results}
\label{sec:results}

\begin{figure}
\caption{Heatmaps for Confusion Matrices obtained from 5-cross validation on our dataset. The reported values, coded in heatmap colors, are the average value between the 5 runs of cross validation. (a) Confusion Matrix for Social platform Classification, (b) Confusion Matrix for upload method classification.}
\label{fig:confusion_matrices}
  \centering
\subfigure[]{
  \label{fig:confmat_social}
	\includegraphics[width=.65\columnwidth] {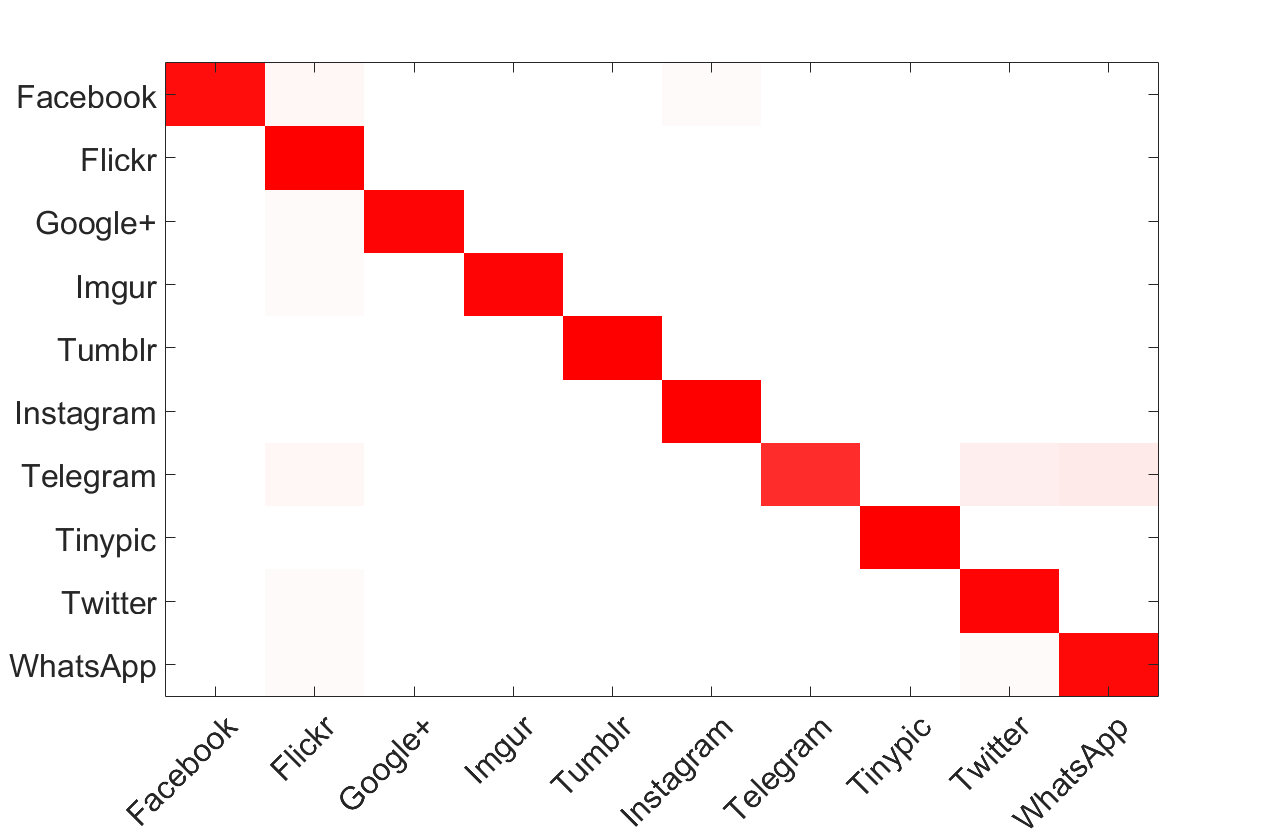}
} 
\subfigure[]{
	\label{fig:confmat_method}
	\includegraphics[width=.3\columnwidth] {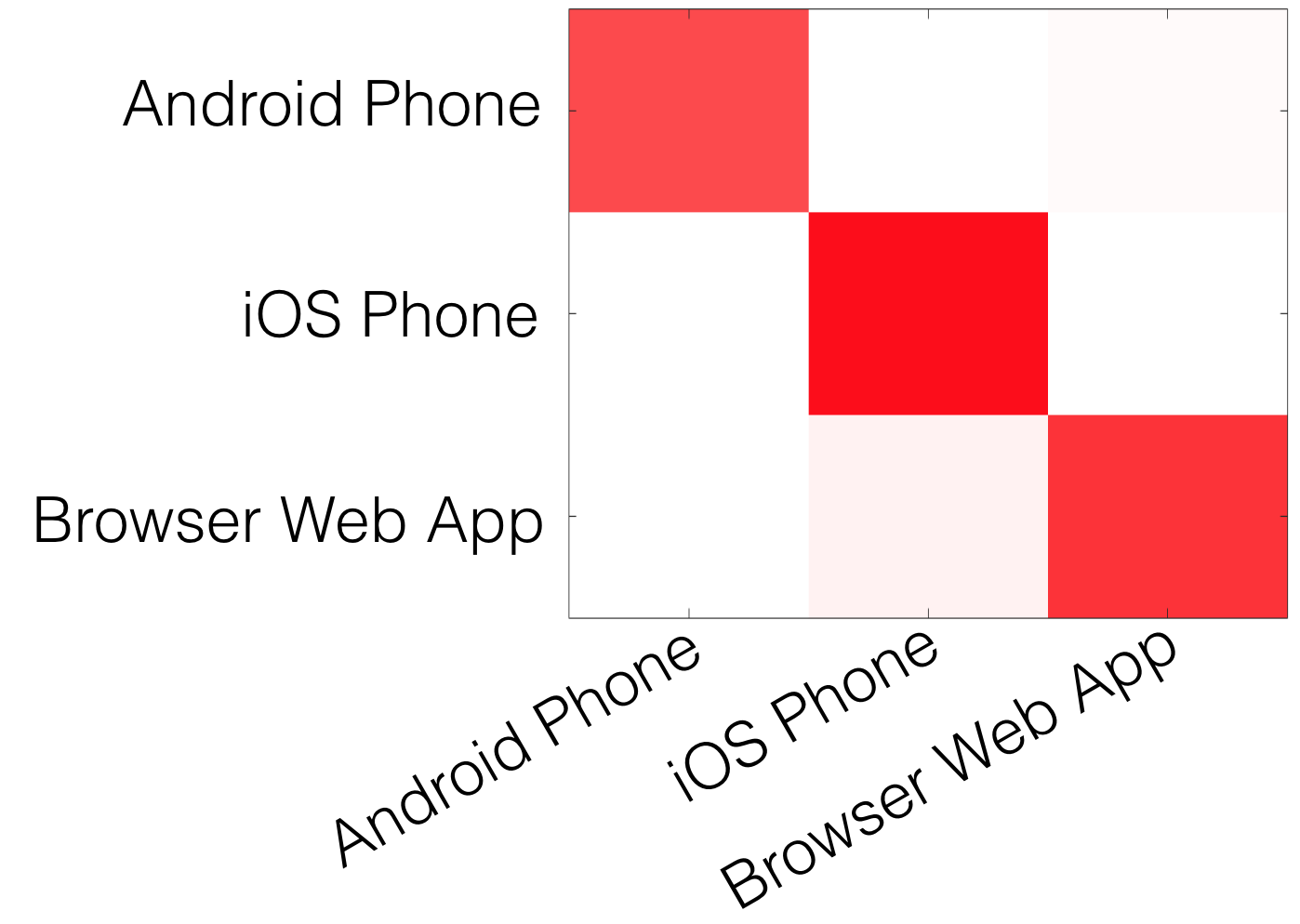}
}
\end{figure}
In this Section, validation results for the proposed approach are reported to demonstrate its goodness. The anomaly detector was validated by taking from our dataset 240 random images that suffered alterations, and 240 images that did not pass through any alteration. The anomaly detector achieved the best error rate, equal to 3.37\%, with K = 3 and T = 2.90.

The entire approach for image ballistics described in the previous Section was then tested through a 5-fold cross validation test. In Figure \ref{fig:confusion_matrices}, confusion matrices reporting the average value through the 5 runs are shown. 

The accuracy obtained for the SNS classification task was 96\% with best K equal to 3 while the accuracy value for the upload client classification task was 97.69\% with an accuracy of 91\% for the prediction of image selection method, given iOS or Android native app as prior. %Accuracy values reported are the average of the 5 cross-validation runs.

A different approach, with a cascade of classifiers, was also tested with each classifier being alternately the predictor for the other one, but the overall results were slightly worse. The classification scheme reported in Figure \ref{fig:classification} was the best approach we obtained throughout our tests.

\subsection{Discussion}
\label{sec:discussion}
In our experiments, we observed that, as happens for different camera devices of the same model \cite{Kornblum2008}, different images, from the same platform, have slightly differences in DQT coefficients. Hence, the most discriminative features were chosen among those listed in Section \ref{sec:ballistics}. At conference time more details about this aspect will be presented.

Another consideration is needed about the SNSs fingerprints described in this work and regarding the fact that all the alterations observed can change according to software development and releases. For these reasons, in order to solve the task of predicting the SNS that processed an image and the client carrying out the upload, the proposed automatic and probabilistic approach is justified for being able to scale and readapt through time, just by updating the reference dataset.

\section{Conclusions and future works}
\label{sec:conclusions}
In this work, we presented a dataset for image ballistic and proposed a classification engine to discover if an image has been processed by a Social Network Service and, if the answer is positive, by which SNS among the 10 considered platforms. The proposed approach performed the task of Image Ballistics with good accuracy by predicting the SNS that process an image and the corresponding upload method, with an accuracy respectively of 96\% and 97.69\%.

We think that this work can open new perspectives on the field of Image Forensics: the approach can be upgraded by considering other formats (e.g., PNG) and new features related to image contents.

% references section

% can use a bibliography generated by BibTeX as a .bbl file
% BibTeX documentation can be easily obtained at:
% http://www.ctan.org/tex-archive/biblio/bibtex/contrib/doc/
% The IEEEtran BibTeX style support page is at:
% http://www.michaelshell.org/tex/ieeetran/bibtex/
%\bibliographystyle{IEEEtran}
% argument is your BibTeX string definitions and bibliography database(s)
%\bibliography{IEEEabrv,../bib/paper}
%
% <OR> manually copy in the resultant .bbl file
% set second argument of \begin to the number of references
% (used to reserve space for the reference number labels box)
%\begin{thebibliography}{1}

%\bibitem{IEEEhowto:kopka}
%H.~Kopka and P.~W. Daly, \emph{A Guide to \LaTeX}, 3rd~ed.\hskip 1em plus
%  0.5em minus 0.4em\relax Harlow, England: Addison-Wesley, 1999.

%\end{thebibliography}
\bibliographystyle{IEEEtran}
\bibliography{bibliography.bib}
%\begin{thebibliography}{}
%\bibitem{kee2011}
%	Kee, E., Johnson, M.K., Farid, H.: Digital image authentication from JPEG headers. IEEE Transactions on Information Forensics and Security, vol. 6, no. 3, pp. 1066–1075, Sep. 2011.

%    \bibitem{djpeg}
%	{{\{DJPEG\} LibJPEG open-source project on GITHUB}} -- {https://github.com/LuaDist/libjpeg}
%\end{thebibliography}

% biography section
% 
% If you have an EPS/PDF photo (graphicx package needed) extra braces are
% needed around the contents of the optional argument to biography to prevent
% the LaTeX parser from getting confused when it sees the complicated
% \includegraphics command within an optional argument. (You could create
% your own custom macro containing the \includegraphics command to make things
% simpler here.)
%\begin{IEEEbiography}[{\includegraphics[width=1in,height=1.25in,clip,keepaspectratio]{mshell}}]{Michael Shell}
% or if you just want to reserve a space for a photo:

%\vfill

% Can be used to pull up biographies so that the bottom of the last one
% is flush with the other column.
%\enlargethispage{-5in}

% that's all folks
\end{document}